\documentclass[aps,twocolumn,pre,eqsecnum]{revtex4}

\usepackage{graphpap}
\usepackage[dvips]{graphicx}
\usepackage[dvips]{graphics}
\usepackage{color}

\begin{document}

\title{Charge Detection in Graphene Quantum Dots}

 \author{J. G\"uttinger, C. Stampfer, S. Hellm\"uller, F. Molitor, T. Ihn and K. Ensslin}

\affiliation{Solid State Physics Laboratory, ETH Zurich, 8093 Zurich, Switzerland}

 
\begin{abstract}
We report measurements on a graphene quantum dot with an integrated graphene charge detector. The quantum dot device consists of a graphene island (diameter $\sim$ 200 nm) connected to source and drain contacts via two narrow graphene constrictions. From Coulomb diamond measurements a charging energy of 4.3 meV is extracted. The charge detector is based on a 45 nm wide graphene nanoribbon placed approx. 60 nm from the island. We show that resonances in the nanoribbon can be used to detect individual charging events on the quantum dot. The charging induced potential change on the quantum dot causes a step-like change of the current in the charge detector. The relative change of the current ranges from 10\% up to 60\% for detecting individual charging events.
\end{abstract}

\maketitle
Graphene~\cite{gei07,kat07}, the two dimensional sp$^2$ allotrope of carbon is a promising material for the development of future nanoelectronics and quantum information processing~\cite{tra07}. This is mainly due to high carrier mobilities~\cite{nov05,zha05} and expected long spin lifetimes. Graphene's gapless electronic structure and predicted Klein tunneling through pn-barriers~\cite{kat06} makes it hard to confine charge carriers by electrostatic means. However, by etching graphene it is possible to make tunable graphene nanodevices, as it has been shown by the fabrication of graphene nanoribbons~\cite{che07,han07,dai08}, interference devices~\cite{mia07,rus08,mol08} and graphene quantum dots~\cite{sta08,sta08b,pon08}. 
In this paper we present an integrated graphene device consisting of a graphene quantum dot with a nearby graphene nanoribbon acting as a quantum-point-contact-like charge detector. Charge detection techniques~\cite{fie93} have been shown to significantly extend the experimental possibilities with quantum dot devices. They are e.g. powerful for detecting spin-qubit states~\cite{elz04,pet05} and molecular states in coupled quantum dots~\cite{dic04}. Furthermore charge detectors have been succesfully used to investigate shot noise on a single electron level and full counting statistics~\cite{gus06}. This makes charge detection highly interesting for advanced investigation of graphene quantum dots and graphene nanosystems in general.

Fig.~1a shows a scanning force microscope image of the all graphene structure. The quantum dot device consists
of two 35 nm wide graphene constrictions separating source (S) and drain (D) contacts from the graphene island 
(diameter $\sim$ 200~nm). 
The constrictions and the island are electrostatically tuned independently by two barrier gates (B1 and B2) and
a plunger gate (PG), respectively. The highly doped silicon back gate (BG)
allows to adjust the overall Fermi level. In addition, we placed a 45 nm wide graphene nanoribbon 60 nm next to the island, which acts as a charge detector (CD), as shown below. 

The sample is fabricated by mechanical exfoliation of natural bulk graphite~\cite{nov04}. 
Single-layer graphene flakes  are transferred to highly doped silicon substrates with a 295~nm silicon oxide top-layer. 
Electron beam lithography (EBL) is used for patterning the isolated graphene flakes by subsequent Ar/O$_2$ reactive ion etching. A second EBL and a lift-off step is performed to place source, drain electrodes and contacts to the lateral gate electrodes (all 2/50~nm Ti/Au). For the detailed fabrication process and the single-layer graphene identification by Raman spectroscopy we refer to Refs.~\cite{sta08,fer06,dav07a}. 

Measurements were performed in a variable temperature insert cryostat at a base temperature of 1.7~K using low-frequency lock-in techniques.

\begin{figure}\centering
\includegraphics[draft=false,keepaspectratio=true,clip,%
                   width=0.98\linewidth]%
                   {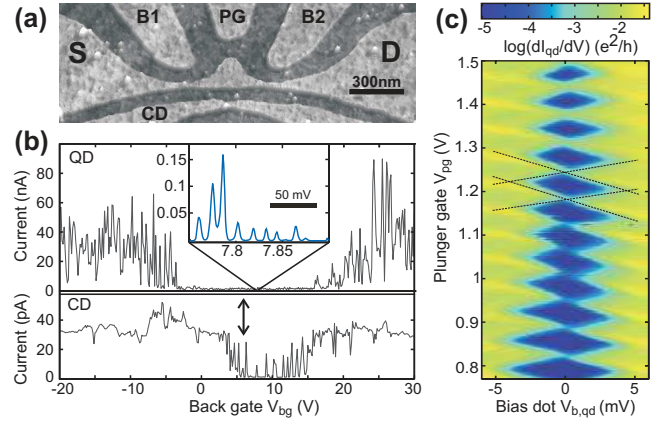}                   
\caption[FIG1]{(Color online)
Nanostructured graphene quantum dot device with nanoribbon and characteristic transport measurements.
(a) Scanning force micrograph of the measured device. The central island is connected to source (S) and drain (D) contacts by two constrictions. The diameter of the dot is 200~nm and the constrictions are 35~nm wide.
The graphene nanoribbon acts as charge detector (CD). Three lateral gates B1, B2 and PG are used to tune the devices.
(b) Back gate characteristics of the quantum dot (QD) device (upper panel) and the charge detector (lower panel), shown in (a). Both measurements were performed at a source-drain (bias) voltage of $V_{b,qd} = V_{b,cd} = 500~\mu$V and at 1.7~K.
Inset shows Coulomb blockade resonances observed inside the transport gap as a function of the back gate voltage over a range of 150~mV. (c) Coulomb blockade diamonds in differential conductance (logarithmic scale) recorded as function of the plunger gate and bias voltage with fixed back gate voltage $V_{bg} = 2$~V. The charging energy is estimated to be $E_c \approx$ 4.3 meV.
} 
\label{trdansport}
\end{figure}

The characterization of the individual devices is shown in Fig.~1b and 1c.
Fig.~1b shows the current $I$ as a function of the back gate voltage at a temperature of~1.7~K of both, the quantum dot device (upper curve) 
and the charge detector (lower curve). In both cases we observe a transport gap~\cite{sta08} extending roughly from -4~V to 15~V and from 4~V to 14~V for 
the quantum dot, and charge detector, respectively. 
From high source-drain voltage ($V_{b,qd}$) measurements (not shown) we estimate the characteristic energy scale
of these effective energy gaps to be about 13 meV, and 8 meV, respectively. This is in reasonable agreement with recent measurements on graphene nanoribbons 
where the transport gap is dominated by the width of the graphene nanostructure~\cite{sol07}.
The large scale current fluctuations are attributed to resonances in the 
graphene constrictions.
By focusing on a smaller back gate voltage range of 150~mV (see inset in Fig. 1b) Coulomb blockade resonances of the quantum dot are resolved
in regions where these resonances are suppressed. In Fig.~1c Coulomb diamond measurements of the quantum dot are shown. The differential conductance of the dot $d I_{qd}/d V_{b,qd}$ is plotted as a function of the bias voltage $V_{b,qd}$ and plunger gate voltage $V_{pg}$ for a fixed back gate  voltage $V_{bg} = 2$~V. From this measurement a charging energy $E_c \approx$ 4.3 meV and a plunger gate
lever arm $\alpha_{pg,qd}$=0.06 is extracted. From further diamond measurements as function of back gate and fixed plunger gate voltage (not shown here) we find a back gate lever arm of $\alpha_{bg,qd}$=0.34.

\begin{figure}\centering
\includegraphics[draft=false,keepaspectratio=true,clip,%
                   width=0.85\linewidth]%
                   {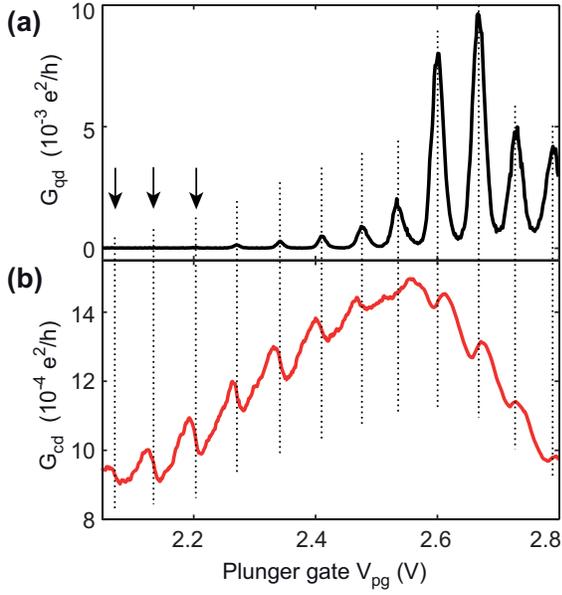}                   
\caption[FIG2]{(Color online). Dot conductance $G_{qd}$ (a) and charge detector conductance $G_{cd}$ (b) as a function of plunger gate voltage $V_{pg}$ for a fixed back gate voltage $V_{bg}$ = 6.5 V. The arrows indicate Coulomb blockade resonances which can be hardly measured by conventional means (a) because the current levels are too low. However, they can be detected by the charge detector (b). Bias on dot: $V_{b,qd} =$ 500 $\mu$V, bias on charge detector: $V_{b,cd}$ = 8.2~mV. 
} 
\label{CD1}
\end{figure}

\begin{figure} \centering
\includegraphics[draft=false,keepaspectratio=true,clip,%
                   width=0.85\linewidth]
                   {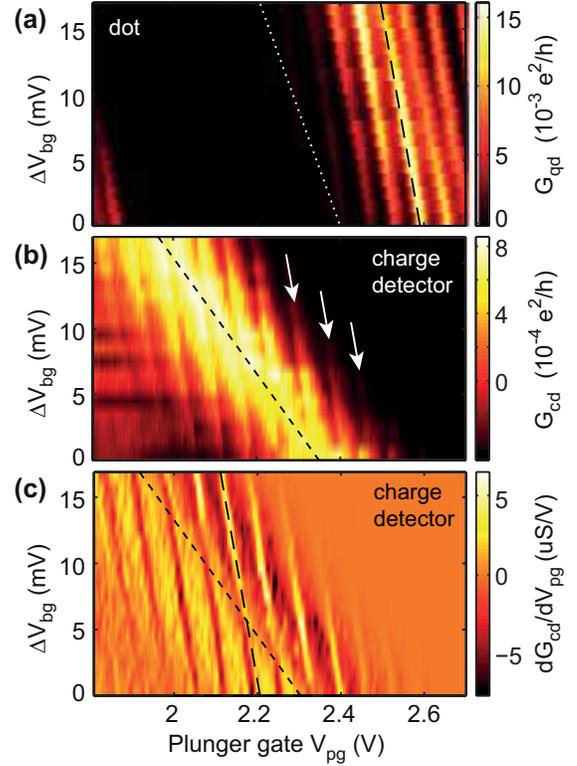}                   
\caption[FIG3]{(Color online). (a) Conductance $G_{qd}$ of the quantum dot as a function of plunger gate voltage $V_{pg}$ and back gate voltage $V_{bg}$. The back gate voltage is converted to a relative scale starting at $V_{bg}$ = 6.505 V with $\Delta V_{bg} = 0$ For this measurement a source drain bias of $V_{b,qd} = 500$~$\mu$V is symmetrically applied. Narrowly spaced periodic lines are Coulomb blockade resonances (black line with long dashes), while the larger scale features are attributed to a modulation of the transmission through the barriers (white dotted line). (b) Simultaneous measurement of the charge detector conductance $G_{cd}$. The broad line with increased conductance is less affected by changing the plunger gate voltage compared to the Coulomb blockade resonances in the dot and it is attributed to a local resonance in the charge detector (short dashed line). In addition to this broad line faint lines with a slope corresponding to the Coulomb blockade resonances in the quantum dot are observed (arrows). (c)  Derivative of the charge detector conductance plotted in (b) with respect to plunger 
gate voltage $dG_{cd}/dV_{pg}$, where the lines with short and long dashes indicate the two different lever arms. 
} 
\label{CD2}
\end{figure}

After having demonstrated the functionality of both devices independently their joint operation is shown in Figs.~2 and 3 where we demonstrate the functionality and high sensitivity of the graphene charge detector.

For these measurements the back gate voltage is set to $V_{bg}=6.5$~V such that
the quantum dot is close to the charge neutrality point 
(see arrows in Fig.~1b) as well as inside the transport gap of the charge
detector. We operate the
charge detector in a regime where strong resonances are accessible, 
in order to make use of steep slopes 
of the conductance as a function of $V_{pg}$
of the order of 4-6 10$^{-4}$~($e^2/h$)/100 mV to detect
individual charging events on the quantum dot.

Fig.~2a shows almost equidistantly spaced ($\Delta V_{pg} = 65 \pm 3.8$~mV) Coulomb blockade resonances 
as function of $V_{pg}$ at $V_{b,qd}=500~\mu$V. 
The strong modulation of the conductance peak amplitudes is due to
superimposed resonances in the graphene constrictions defining the 
island~\cite{sta08}. In Fig.~2b we plot the simultaneously measured conductance through the nearby charge detector
(at a bias voltage of $V_{b,cd}=8.2$~mV) for the same $V_{pg}$ range.
On top of the peak-shaped CD resonance we observe conductance steps which are 
well aligned (see dotted lines) with single charging events on the nearby quantum dot.

Figs.~3a and 3b show 2D plots of a set of traces corresponding to those shown
in Figs.~2a and~2b, taken for different back gate voltages and $V_{b,cd}=V_{b,qd}=500$~$\mu$V.
Fig.~5a shows Coulomb blockade resonances in the quantum dot conductance
following a relativ lever arm of PG and BG of $\alpha_{pg,qd}/\alpha_{bg,qd}$ = 0.18 
(see black dashed line). In Fig.~3b these
resonances are observed through charge detection and are marked with arrows. The charge detector
resonance used for detection can be distinguished from the dot resonances by its larger width and
its different slope given by $\alpha_{pg,cd}/\alpha_{bg,cd}$ = 0.04 (black dashed line). 
This reduced slope is due to the larger distance of the CD nanoribbon to the plunger 
gate ($\sim$ 350~nm) as compared to the island-plunger gate distance.
The modulation of the Coulomb blockade resonances in Fig.~3a is due to
resonances in the tunneling constriction  
located around 300~nm away from the PG (see Fig.~1a). This leads to a slope of 0.08 for these
peak modulations (see white dotted line).
Independent of this modulation we identify single charging events on the quantum dot as
conductance fringes (see arrows in Fig.~3b)
on top of the up and down slope
of the CD resonance.
This can even better be seen by numerical differentiation of $G_{cd}$ vs.
$V_{pg}$, as shown in Fig.~3c. Here the sharp conductance changes due to the charging events in the dot are strongly pronounced, and both relative lever arms to the Coulomb blockade peaks and
the constriction resonance in the charge detector are indicated by dashed lines. The detection range can be improved by increasing the bias $V_{b,cd}$, leading to a broadening of  
the constriction resonance, as seen by comparing Fig.~3c with Fig.~2b. 

From the measurement shown in Fig.~2b a nanoribbon conductance change of up to 10\% can be exctracted for a single charging event. For lower bias voltages (e.g. $V_{b,cd} = 500~\mu$V) the change in the conductance can be incresed to 60\%.

In conclusion, we demonstrated the functionality of an integrated graphene charge detector based on a nanoribbon nearby a graphene quantum dot. We confirm the detection of charging events in regimes where Coulomb blockade resonances can hardly be measured (i.e. resolved) because the current levels are too low (see, e.g., arrows in Fig.~2a). In contrast to state-of-the-art quantum point contact charge detectors we do not make use of slopes to quantized conductance plateaus. We rather use local resonances in the graphene nanoribbon to detect charging. We anticipate, that this technique will play an important role for the investigation of the electron-hole crossover and spin-states in graphene quantum dots.\\

The authors wish to thank R.~Leturcq, P.~Studerus, C.~Barengo, P.~Strasser, A.~Castro-Neto and K.~S.~Novoselov for helpful discussions. 
Support by the ETH FIRST Lab, the Swiss National Science Foundation and NCCR nanoscience are gratefully acknowledged.

\newpage


\begin{thebibliography}{99}


\bibitem{gei07}
A. K. Geim and K. S. Novoselov, Nat. Mater.~{\bf 6}, 18 (2007).

\bibitem{kat07}
M. I. Katsnelson, Mater. Today~{\bf 10}, 20 (2007).


\bibitem{tra07}
B. Trauzettel, D.V. Bulaev, D.~Loss, and G.~Burkard, Nature Physics,~{\bf 3}, 192, (2007).

\bibitem{nov05}
K. S. Novoselov, A. K. Geim, S. V. Morozov, D.~Jiang, M.~I.~Katsnelson, I.~V.~Grigorieva, S.~V.~Dubonos, and A. A. Firsov,  Nature,~{\bf 438}, 197-200, (2005).

\bibitem{zha05}
Y. Zhang, Y.-W.~Tan, H.~L.~Stormer, and P.~Kim, Nature,~{\bf 438}, 201-204, (2005).

\bibitem{kat06}
M. I. Katsnelson, K. S. Novoselov, amd A. K. Geim, Nature Phys.~{\bf 2},
620–625 (2006).





\bibitem{che07}
Z. Chen, Y. Lin, M. Rooks, and P. Avouris, Physica E,~{\bf 40}, 228, (2007).

\bibitem{han07}
M. Y. Han, B. \"Ozyilmaz, Y. Zhang, and P. Kim, Phys. Rev. Lett.,~{\bf 98}, 206805 (2007).

\bibitem{dai08}
X. Li, X. Wang, L. Zhang, S. Lee, and H. Dai, Science,~{\bf 319}, 1229 (2008).

\bibitem{mia07}
F.~Miao, S.~Wijeratne, Y.~Zhang, U.~C.~Coskun, W.~Bao, and C.~N.~Lau, Science,~{\bf 317}, 1530 (2007).

\bibitem{rus08}
S. Russo, J. B. Oostinga, D. Wehenkel, H. B. Heersche, S. S. Sobhani,
L. M. K. Vandersypen, and A. F. Morpurgo, Phys. Rev. B,~{\bf 77}, 085413 (2008).

\bibitem{mol08}
F. Molitor, A. Jacobson et.al., to be published (2008)
\bibitem{sta08}
C. Stampfer, J. G\"uttinger, F. Molitor, D. Graf, T. Ihn, and K. Ensslin, Appl. Phys. Lett.,~{\bf 92}, 012102 (2008).

\bibitem{sta08b}
C. Stampfer, E. Schurtenberger, F. Molitor, J. Güttinger, T. Ihn, and K. Ensslin,
Nano Lett.~{\bf 8}(8), 2378 (2008).

\bibitem{pon08}
L. A. Ponomarenko, F. Schedin, M. I. Katsnelson, R.~Yang, E.~H.~Hill, K.~S.~Novoselov, and A.~K.~Geim,
Science,~{\bf 320}, 356 (2008).



\bibitem{fie93}
M. Field, C. G. Smith, M. Pepper, D. A. Ritchie, J. E. F. Frost, G. A. C. Jones, and D. G. Hasko,
Phys. Rev. Lett.~{\bf 70}, 1311 (1993).

\bibitem{pet05}
J. R. Petta, A. C. Johnson, J. M. Taylor, E. A. Laird,
A. Yacoby, M. D. Lukin, C. M. Marcus, M. P. Hanson,
and A. C. Gossard, Science,~{\bf 309}, 2180-2184 (2005).

\bibitem{elz04}
J. M. Elzerman, R. Hanson, L. H. Willems van Beveren, B. Witkamp, L. M. K. Vandersypen, and L. P. Kouwenhoven, Nature~{\bf 430}, 431 (2004).
 
\bibitem{dic04}
L. DiCarlo, H. Lynch, A. Johnson, L. Childress, K. Crockett, C. Marcus, M. Hanson, and A.
Gossard, Phys. Rev. Lett.~{\bf 92}, 226801 (2004) .


\bibitem{gus06}
S. Gustavsson, R. Leturcq, B. Simovic, R. Schleser, T. Ihn, P. Studerus, and K. Ensslin, Phys. Rev. Lett.~{\bf 96}, 076605 (2006).
















\bibitem{nov04}
K. S. Novoselov, A. K. Geim, S. V. Morozov, D.~Jiang, M.~I.~Katsnelson, S.~V.~Dubonos, I.~V.~Grigorieva, and A. A. Firsov, Science,~{\bf 306}, 666, (2004).

\bibitem{fer06} 
A.~C. Ferrari, J.~C. Meyer, V. Scardaci, C. Casiraghi, M. Lazzeri, F. Mauri,
S. Piscanec, D. Jiang, K.~S. Novoselov, S. Roth, and A.~K. Geim,
Phys. Rev. Lett. {\bf 97}, 187401 (2006).

\bibitem{dav07a} D. Graf, F. Molitor, K. Ensslin, C. Stampfer, A. Jungen, C. Hierold, and
L. Wirtz, Nano Lett., {\bf 7}, 238 (2007).

\bibitem{sol07}
F.~Sols, F.~Guinea, and A.~H.~Castro Neto, Phys. Rev. Lett.,~{\bf 99}, 166803 (2007).




















\end{thebibliography}
\end{document}